\documentclass{desyproc}

\begin{document}
\title{Intermediate Scale Accidental Axion and ALPs}

\author{{\slshape Alex G. Dias}\\[1ex]
Universidade Federal do ABC -- Centro de Ci\^encias Naturais e Humanas, Santo Andr\'e, Brasil
}

\contribID{familyname\_firstname}

\acronym{Patras 2014} 
\doi  

\maketitle

\begin{abstract}
We discuss the problem of constructing models containing an axion and  axion-like particles, motivated by astrophysical observations, with decay constants at the intermediate scale ranging from $10^9$GeV to $10^{13}$GeV. We present  examples in which the axion and axion-like particles arise accidentally as pseudo Nambu-Goldstone bosons of automatic global chiral symmetries, in models having exact discrete symmetries.
\end{abstract}


Pseudoscalar bosons very weakly interacting  and having low mass --  below the eV scale for example -- are common in the particle spectra of theories aiming to answer questions left open by the Standard Model (SM) like, for example, the  CP conservation of the strong interactions, and the nature of dark matter.  The axion is a prominent example occurring in extensions of the SM  constructed to solve the strong CP problem \cite{Peccei:1977hh,Weinberg:1977ma,Wilczek:1977pj}. Not connected with this last problem, but with interactions similar to the axion, are the axion-like particles (ALPs). For a review  see~\cite{Jaeckel:2010ni}.

Axions and ALPs arise in models containing global symmetries that, besides being spontaneously broken, are also explicitly broken conferring small masses for those particles characterizing them as pseudo Nambu-Goldstone bosons. There are  astrophysical phenomena motivating the construction of  models containing at last one ALP  in addition to the axion~\cite{Dias:2014osa}. Thus, we consider first a simple case of  SM extensions containing two global chiral  symmetries, U(1)$_1$ $\times$ U(1)$_2$ with each factor broken spontaneously by vacuum expectation values (vev) $\langle\sigma_i\rangle=v_i/\sqrt2$ of SM singlets complex $\sigma_i(x)=[v_i+\rho_i(x)]e^{i\,a^\prime_i(x)/f_{a_i^\prime}}/\sqrt2$, $i=1,\,2$, where the decay constants $f_{a_i^\prime}$ depend on $v_i$ and the vev of other scalar fields  carrying charge of the  U(1)$_{is}$. At energies much below the  scales $v_{i}\sim f_{a_i^\prime}$, the low energy effective Lagrangian contains the $a^\prime_i(x)$ fields interactions with the gluons and electromagnetic field, through the field strengths $G_{\mu\nu}$ and $F_{\mu\nu}$,
\begin{equation}
{\mathcal{L}} \supset\frac{1}{2}\, \sum_{i=1}^{2}\partial_\mu a_i^\prime\, \partial^\mu a_i^\prime
- \frac{\alpha_s}{8\pi} \sum_{i=1}^{2} C_{ig} \frac{a_i^\prime}{f_{a_i^\prime}}\,   G_{\mu\nu}^a \tilde{G}^{a,\mu\nu}
 - \frac{\alpha}{8\pi}
\sum_{i=1}^{2}C_{i\gamma} \frac{a_i^\prime}{f_{a_i^\prime}} \, F_{\mu\nu} \tilde{F}^{\mu\nu}\,.
\label{ALP_leff}
\end{equation}
We have omitted the $a^\prime_i(x)$ interactions with fermionic fields since we are not going to  discuss them here.
The anomaly coefficients, $C_{ig}$ and $C_{i\gamma}$ are model dependent but  typically of order one in common models~\cite{Dias:2014osa}. It is observed in Eq. (\ref{ALP_leff}) that the fields $a^\prime_i(x)$ can be made very weakly interacting if the decay constants $f_{a_i^\prime}$ are sufficiently high -- much above the electroweak scale  $\sim$ 246 GeV. Equation (\ref{ALP_leff}) has the axion as the particle excitation of the field $A(x)= C_{1g} \,{a_1^\prime(x)}f_A/{f_{a_1^\prime}}+ C_{2g}\,{a_2^\prime(x)}f_A/{f_{a_2^\prime}}$, with the ALP the one of the field $a(x)=C_{2g} \,{a_1^\prime(x)}f_a/{f_{a_2^\prime}}$ - $C_{1g}\,{a_2^\prime(x)}f_a/{f_{a_2^\prime}}$, and the decay constants $1/f_A^2=1/f_a^2=(C_{1g}/f_{a_1^\prime})^2+(C_{2g}/f_{a_2^\prime})^2$. The axion and the ALP couplings with the photon  are defined through $-\frac{g_{\phi\gamma}}{4}\,\phi\, F_{\mu\nu}\widetilde{F}^{\mu\nu}$, where $\phi=A,\,a$, from Eq. (\ref{ALP_leff}) as $g_{A\gamma}=\frac{\alpha\,f_A}{2\pi }\left( C_{1g}C_{1\gamma}/f_{a_2^\prime}+  C_{2g}C_{2\gamma}/f_{a_2^\prime}
- 1.95\right)$ and $g_{a\gamma}=\frac{\alpha\,f_A}{2\pi}\left( C_{1g}C_{2\gamma} - C_{1\gamma}C_{2g} \right)/f_{a_1^\prime} f_{a_2^\prime}$.
The number -1.95 in the above is a universal contribution from the  axion-neutral pion  mixing.  $g_{A\gamma}$ and $g_{a\gamma}$ are the main  couplings of the axion and ALPs, respectively, giving rise to notable phenomena like photon-axion/ALP oscillations~\cite{Jaeckel:2010ni}. Intermediate e\-ner\-gy scales for the decay constants such that $10^9$GeV$\lesssim f_{a_i^\prime}\lesssim $ $10^{13}$GeV are specially in\-te\-res\-ting. This furnishes the values of $g_{A\gamma}$ and $g_{a\gamma}$  within the ranges to be probed directly in new experiments, required  to potentially explain some phenomena hinted by astrophysical observations. Depending on their masses,  the axion and the ALP could also be cold dark matter candidates.

The pseudo-Nambu-Goldstone field $A(x)$ is associated to the Peccei-Quinn symmetry, U(1)$_{PQ}$ ($\subset$ U(1)$_1$ $\times$ U(1)$_2$), which is a chiral global symmetry with the special property of being anomalous -- explicitly broken by non-perturbative effects -- in the quarks sector leading to the interaction term $\frac{1}{f_A}A\,G_{\mu\nu}^a \tilde{G}^{a,\mu\nu}$. This allows the elimination of the CP violation term $\bar\theta\, G_{\mu\nu}^a {\tilde G}^{a,\mu\nu}$ absorbing the $\bar\theta$-parameter into the axion field as  $A+\bar\theta\,f_A\rightarrow A$  \cite{Peccei:1977hh}. Also, it is generated a potential $V(A) \simeq m_A^2 f_A^2\left[1-\cos\left(\frac{A}{f_A}\right)\right]$, in which the axion mass is $m_A=\frac{m_\pi f_\pi}{f_A}\frac{\sqrt{z}}{1+z}\simeq {6\,  {\rm meV}}\times\left(\frac{10^{9}\, {\rm GeV}}{f_A}\right)$, with $m_\pi$ the mass of the pion,  $f_\pi$ its decay constant, and $z\approx0.56$~\cite{Weinberg:1977ma}.
$V(A)$  leads to the result that the effective CP violation parameter turns out to be zero by the fact that $\langle \frac{A}{f_A}\rangle=\theta_{eff}=0$. This solves the strong CP problem, which is a fine tuning problem once $\bar\theta$ must be very small, arising due the non-observation of a electric dipole moment of the neutron, whose actual measurements limits $|\bar\theta|\lesssim10^{-10}$ \cite{Baker:2006ts}.

Differently from the axion the ALP remains massless -- as Nambu-Goldstone boson of the combined U(1)$_i$ out of U(1)$_{PQ}$ -- unless there are additional interactions explicitly breaking its associate global symmetry and, thus, generating an extra potential $\delta V (a^\prime_i)$. Massive ALPs  have been implemented in ultra-violet completions of the SM for several purposes~\cite{Dias:2014osa}. But the explicit breakdown of the global symmetries must occur in a controlled way to get  appropriate ALPs masses and preserve the solution of the strong CP problem. In fact, it is not expected that gravitational interactions conserve global symmetries. It is known that operators suppressed by the Planck scale $M_{\rm Pl}$ like $\sigma_1^n \sigma_2^k/M_{\rm Pl}^{D-4}$, with $D=n+k>4$, might bring corrections to the axion potential such that $\delta\bar\theta>10^{-10}$   if not forbidden until a certain dimension $D \gtrsim {9}/\left[{1-0.1\,{\rm log}\left({f_A/10^9{\rm GeV}}\right)}\right]$  \cite{Ghigna:1992iv}. Discrete gauge symmetries $Z_N$ \cite{Krauss:1988zc}, with $N=D+1$, have been used to resolve the problem  of having dangerous effective operators  in axion models~\cite{Dias:2002hz}, and also  in models containing axion plus ALPs \cite{Dias:2014osa}. Another compelling reason for these $Z_N$ symmetries is that if they are appropriately postulate  anomalous  symmetries like U(1)$_{PQ}$ may arise as automatic  quasi exact symmetries. This avoids the non-natural  impositions of global continuous symmetries which are already explicitly broken, as is the case of the U(1)$_{PQ}$ symmetry.

A model in which the axion and ALPs have their masses and couplings controlled by discrete symmetries is the ${\mathbb{Z}}_{13}\otimes {\mathbb{Z}}_5\otimes{\mathbb{Z}}_{5}^\prime$ model proposed in \cite{Dias:2014osa}. It is a hybrid of invisible axion models~\cite{Kim:1979if}  plus an ALP. The field content of the model beyond the SM one is: four SM Higgs doublets $H_b$, an SU(2)$_L$ triplet $T$, a vector-like color triplet $(Q_L,Q_R)$, three right-handed neutrinos $N_{iR}$, and two SM singlet fields $\sigma_{1,2}$. The fields charges of the imposed discrete symmetry ${\mathbb{Z}}_{13}\otimes{\mathbb{Z}}_5\otimes{\mathbb{Z}}^\prime_5$  are shown in Table \ref{tableZns}. It can be shown that both the Yukawa interaction terms and the renormalizable scalar potential of this model have two accidental global chiral symmetries and, thus, an axion and one ALP.  Additionally, the interaction terms $\overline{L}_{i}\widetilde{H}_N N_{j R}$ and $\overline{(N_{iR})^c} \sigma_1 N_{jR}$, allowed by the discrete symmetries of the model   lead to a seesaw mechanism for generating small masses to the active neutrinos.

\begin{table}[h]
\[
\begin{array}{|c|cccccccccccccccc|}
\hline
\psi_i & q_L & u_R & d_R & L & N_{R} & l_R & H_u & H_d & H_l & H_N &
\sigma_2 & T  & Q_L & Q_R & \sigma_1 & \\
\hline
{\mathbb{Z}}_{13} & \omega^5_{13} & \omega^3_{13}  & \omega^{8}_{13} & \omega^{9}_{13} & \omega^3_{13} &
\omega^{7}_{13} & \omega^{11}_{13} &  \omega^{10}_{13} &  \omega^2_{13} &  \omega^{7}_{13} &
\omega^{12}_{13} & \omega^{9}_{13} &  1 & \omega^6_{13} &\omega^{7}_{13}&
\\[.5ex]
\hline
{\mathbb{Z}}_5 & 1 &  \omega_5 &  \omega_5^{4} &  1 &
\omega_5 &  \omega^{4}_5 &  \omega_5  &
\omega_5   & \omega_5  & \omega_5 & 1 &
\omega_5^2 & \omega_5 & \omega^{3}_5 &
\omega^3_5 &
\\[.5ex]
\hline
{\mathbb{Z}}_{5}^{\prime } & 1 & \omega^{4}_5  & 1 & 1 & \omega^{2}_5 &
\omega^{4}_5 & \omega^{4}_5 &  1 &  \omega_5  & \omega^{2}_5 & \omega_5 & \omega^{3}_5 &  1 &
\omega^{4}_5 & \omega_5 & \\[.5ex]
\hline
\end{array}
\]
\caption{\label{tableZns}
${\mathbb{Z}}_N$ charges  of the
${\mathbb{Z}}_{13}\otimes{\mathbb{Z}}_5\otimes{\mathbb{Z}}^\prime_5$ model, where $\omega_{13}\equiv e^{i2\pi/13}$
and $\omega_5\equiv e^{i2\pi/5}$.
}
\end{table}

The lowest dimensional effective operator invariant by the discrete symmetry, but breaking U(1)$_1$$\otimes$U(1)$_2$ explicitly is $\frac{g}{M_{\rm Pl}^{10}}\,H_N^\dag H_d\sigma_1^{*5}\sigma_2^7$. This generates a tiny mass  $m_{a}\sim10^{-33}$ eV to the ALP,  assuming the  values $v_1\simeq f_A\simeq 10^{10}$ GeV,  $v_1\simeq 7.5\times10^{10}$ GeV, taking the Higgses doublets vevs $v_W\simeq 100$ GeV, the Planck scale $M_{\rm Pl}=10^{19}$ GeV, and $g=1$. The axion mass is approximately $m_a\simeq 0.6$ meV, not been affected in a significantly manner by the Planck suppressed effective operators. For the couplings $g_{A\gamma}$, $g_{a\gamma}$, the coefficients entering in them are $C_{1g}=1$, $C_{2g}=3$, $C_{1\gamma}=6$, $C_{2\gamma}=4$, where we assume the electric charge of $Q_{L,R}$ equal to one. This furnishes $|g_{A\gamma}|\simeq 4\times 10^{-13}$ GeV$^{-1}$, and $|g_{a\gamma}|\simeq 2\times10^{-13}$ GeV$^{-1}$. With these values for the coupling to photons and mass the axion can be a dark matter candidate, but still outside the region to be probed directly by present experiments. The ALP in this model has the coupling to photons and mass  within the range required  explanation for the soft X-ray excess from Coma cluster \cite{Conlon:2013txa}, and also being in the reach of proposed experiments \cite{Dias:2014osa}.

We present another construction containing, in addition to the axion, two photophilic ALPs is the $Z_{11}\otimes Z_{9}\otimes Z_{7}$ model. It is motivated by distinct ranges of coupling to photons and mass required to explain the anomalous transparency of the Universe~\cite{Meyer:2013pny} (see \cite{Dias:2014osa} and references therein), and the unidentified X-ray line of 3.55 keV  found in recent observations \cite{Bulbul:2014sua}. With the latter  supposed to be due a two photon decay  of a dark matter ALP with mass of 7.1 keV \cite{Higaki:2014zua}. The field content of the $Z_{11}\otimes Z_{9}\otimes Z_{7}$ model has beyond the SM fields: one vector-like color triplet $(Q_L,Q_R)$; two noncolored vectorial
charged fermions $\left(E_{L},\, E_{R}\right)$, $\left(E^\prime_{L},\, E^\prime_{R}\right)$; three right-handed neutrinos $N_{iR}$;  and three SM singlet fields $\sigma_{i}$. The fields charges of the discrete symmetry are shown in Table \ref{tab:irreps2}.

\begin{table}[ht]\centering
\vspace{-\baselineskip}
\begin{equation}
\nonumber
\begin{array}{|c|c|c|c|c|c|c|c|c|c|c|c|c|c|c|c|c|}
\hline\rule[0cm]{0cm}{.9em}
& q_L & u_R & d_R & L & l_R & N_R & H &  Q_L & Q_R & \sigma_1 & \sigma_2& E_L & E_R & \sigma_3 & E^\prime_L & E^\prime_R
\\[-.1ex]
\hline\rule[0cm]{0cm}{1em}
{\mathbb{Z}}_7 & 1 & \omega_7^3 & \omega_7^4 & 1 & \omega_7^4 & \omega_7^3 & \omega_7^3 & \omega_7^5 &
    \omega_7^3 & \omega_7 & 1 & \omega_7^5  & \omega_7^4 & \omega_7^1 & \omega_7^5 & \omega_7^3
\\[-.1ex]
\hline\rule[0cm]{0cm}{1em}
{\mathbb{Z}}_9 & 1 & \omega_9^5 & \omega_9^4 & \omega_9^6 & \omega_9 & \omega_9^2 & \omega_9^5 & 1 &
    \omega_9^8 & \omega_9^5 & \omega_9 & \omega_9^6  & 1 & 1 & \omega_9^1  & \omega_9^5
\\[-.1ex]
\hline\rule[0cm]{0cm}{1em}
{\mathbb{Z}}_{11} & 1 & \omega_{11}^3 & \omega_{11}^8 & \omega_{11}^{2} & \omega_{11}^{10} & \omega_{11}^5 &
\omega_{11}^3 & \omega_{11}^9 & \omega_{11}^7 & \omega_{11} & 1 &1 & \omega_{11}^{10} & 1 &  \omega_{11}^{10}  & \omega_{11}^{9}
\\[-.1ex] \hline
\end{array}
\end{equation}
\caption{${\mathbb{Z}}_N$ charges  of the ${\mathbb{Z}}_7\otimes{\mathbb{Z}}_9\otimes{\mathbb{Z}}_{11}$ model.}
 \label{tab:irreps2}
\end{table}

It is shown in \cite{Dias:2014osa} that the  Yukawa interaction terms and the renormalizable scalar potential allowed by the discrete symmetry have three global chiral symmetries, U(1)$_1\otimes$U(1)$_2\otimes$U(1)$_3$. Each one of them is spontaneously broken by the respective singlet vevs $\langle\sigma_i\rangle=v_1/\sqrt2$, with the axion being related to  related to U(1)$_1$, the ALP $a_2$ to U(1)$_2$, and the ALP $a_3$ to U(1)$_3$.  There is no one relevant Planck suppressed operator correcting the axion mass. The lowest dimension operators breaking U(1)$_2$ and U(1)$_3$ are $\frac{g}{M_{\rm Pl}^{5}}\,(\sigma_2)^9$ and $\frac{g^\prime}{M_{\rm Pl}^{3}}\,(\sigma_2)^7$, respectively. In order to give an example we set  $v_2= 10^9$ GeV,  $v_3=3\times 10^9$ GeV, $g=1$, and $g^\prime\approx 0.1$. This furnishes $m_{a_2}\approx  10^{-7}$ eV, $g_{a_2\gamma}\approx2.2\times10^{-11}$ GeV$^{-1}$  to the ALP $a_2$ making it able to explain the anomalous transparency of the Universe~\cite{Meyer:2013pny}, and also in the search range of the experiment ALPS II~\cite{Bahre:2013ywa}; and $m_{a_3}\approx$ 7.1 keV, $g_{a_3\gamma}\approx7.7\times10^{-13}$ GeV$^{-1}$  to the ALP $a_3$ so that it can explain the 3.55 keV line  through its two photons decay~\cite{Higaki:2014zua}.

For other developments on discrete symmetries  originating from string theory see \cite{Kim:2014ffa}.



\textbf{Acknowledgments --} This work is  partially supported by CNPq (grant 303094/2013-3)  and FAPESP (grant  2013/22079-8). The author thanks the support of the workshop organizers.


\begin{footnotesize}

\end{footnotesize}


\end{document}